\documentclass{sf2a-conf2010}
\usepackage{graphicx}
\usepackage{hyperref}

\begin{document}
\TitreGlobal{SF2A 2010}
%
\title{Multiple stellar populations in Galactic globular
  clusters: observational evidence}
\author{A.\ P.\ Milone}\address{Dipartimento  di   Astronomia,  Universit\`a  di
  Padova, Vicolo dell'Osservatorio 3, Padova, I-35122, Italy}
\author{G.\ Piotto}\address{Dipartimento  di   Astronomia,  Universit\`a  di
  Padova, Vicolo dell'Osservatorio 3, Padova, I-35122, Italy}
\author{L.\ R.\ Bedin}\address{Space Telescope Science Institute, 3700
  San Martin Drive, Baltimore, MD 21218, USA}
\author{A.\ Bellini}\address{Dipartimento  di   Astronomia,  Universit\`a  di
  Padova, Vicolo dell'Osservatorio 3, Padova, I-35122, Italy}
\author{A.\ F.\ Marino}\address{Max Plank Institute for Astrophysics,
  Postfach 1317, 85741, Garching, Germany }
\author{Y.\ Momany}\address{European Southern Observatory, Alonso de Cordova 3107, Vitacura, Santiago, Chile }
\runningtitle{Multiple populations in globular clusters }
%
\setcounter{page}{237}
\index{MILONE, A. }
\index{PIOTTO, G. }
\maketitle
\begin{abstract}
An increasing number of both photometric and spectroscopic observations 
over the last years have shown the existence of distinct sub-populations in many Galactic globular clusters
and shattered the paradigm of globulars hosting single, simple stellar populations.
 These multiple populations manifest themselves in a split of different evolutionary sequences in the cluster color-magnitude diagrams and in star-to-star abundance variations. In this paper we will summarize the observational scenario.
  
\end{abstract}
\begin{keywords}
Stars: abundances, Stars: atmospheres,  Stars: population II, Galaxy:
globular clusters
\end{keywords}
\section{Introduction}
\label{intro}
In recent years
an increasing amount of photometric and spectroscopic observational
evidence have shattered the paradigm of globulars as the prototype
of single, simple stellar populations (see Piotto\ 2009 for a recent
review).   Spectroscopic studies have demonstrated that most globular
clusters (GC) have no detectable spread in their iron content
 and also $s$-process elements do not exhibit large star-to-star
 variations in the majority of globulars  (e.g. Carretta et al.\ 2009a
 and references therein). 
On the contrary, every time we have at our disposal a large sample of
stars for a given GC, 
 star-to-star variations in the light elements C, N,
O, Na, and Al have been clearly detected
 (e.g. Carretta et al.\ 2009b, Pancino et  al.\ 2010 and references therein).
These variations are related to correlations and anticorrelations,
which indicate the occurrence of high temperature hydrogen-burning
processes (including CNO, NeNa, MgAl cycles) and cannot occur in
presently observed low mass GC stars.    

Today it is widely accepted that the observed light-elements variations
provide strong support to the presence of multiple stellar populations
in GCs with the second generations formed from the material polluted
by a first generation of stars. On the contrary the debate on the
nature of possible polluters is still open (e.g. D'Antona et
al.\ 2004, Decressin et al.\ 2007).

While  abundance variations are well known since the early
sixties, it was only the recent spectacular discovery of multiple
sequences in the color-magnitude diagram (CMD) of several GCs that
provides an un-controversial 
prove of the presence of multiple stellar populations in GCs and
brought new interest and excitement in GCs research (e.g. Piotto et
al.\ 2007).
Photometric clues, often easy to detect simply by the inspection of
high-accuracy CMDs, arise in form of multiple main sequences (MS,
Bedin et al.\ 2004, Piotto et al.\ 2007, Milone et al.\ 2010), split
sub-giant branch (SGB, Milone et al.\ 2008, Anderson et al.\ 2009,
Piotto \ 2009), and multiple red-giant branch (RGB, Marino et
al.\ 2008, Yong et al.\ 2008, Lee et al.\ 2009).

Many population properties, like the
chemical composition, the spatial distribution, the fraction of stars in
each population and their location in the CMD apparently differ from
cluster to cluster. 
Multiple stellar populations have been detected for the first time in
the Milky Way satellite
 $\omega$ Centauri in form of either multiple MSs
(e.g. Anderson\ 1997, Bedin et al.\ 2004, Bellini et al.\ 2010), 
multiple SGBs (e.g. Sollima et al.\ 2005), multiple RGBs (Lee et
al.\ 1999, Pancino et al.\ 2000) and large star-to-star variation in
iron and $s$-elements (e.g. Johnson et al.\ 2010, Marino et
al.\ 2010). Due to its large mass this GCs have been always considered
as a peculiar stellar system and often associated to the remnant of a
dwarf galaxy.

The `extreme' case of $\omega$ Centauri is not analyzed in this
work where we focus on `normal' GCs.   
The following sections are an attempt to define
some groups of `normal' clusters that share similar properties.

\section{Light-elements correlations and spread RGB. The case of NGC 6121.}
\label{NGC6121}
The Na-O anticorrelation has been 
observed in all the Galactic GCs studied to date and indicates the
presence of multiple stellar populations in GCs. 
These chemical inhomogeneities revealed themselves a spread or bimodal
distribution of stars along the RGB when sensitive colors are used.
%

In this context the nearby GC NGC 6121 (M4) is one of the most studied cases
(e.g. Ivans et al.\ 1999, Marino et al.\ 2008).
An high resolution spectroscopic study conduced on a
sample of more than a hundred RGB stars in this GC, has recently
demonstrated that 
two different stellar populations are present (Marino et
al.\ 2008). 
These two populations have been identified by means of a strong 
dichotomy in the sodium abundance, that is related to a bimodality in
the CN band 
strengths. The presence of two different groups of stars is visible
also on the Na-O anticorrelation: a bulk of stars with Na and O
resembling the halo field content, and likely associated to the first stellar
generation, can be easily distinguished from stars enhanced in Na and
with lower O abundance, associated with the second generation.
The two populations of Na-poor/O-rich/CN-weak and
Na-rich/O-poor/CN-strong stars populate two different regions
along the RGB, when plotted in a {\it U} vs. ({\it (U-B)}) CMD. The Na-rich
group defines a narrower sequence on the 
red side of the mail RGB locus, while the Na-poor one populates a bluer, more
spread portion of the RGB. 
The RGB spread is visible from the base of the RGB to its tip, and
it is due to the dichotomy in the CN bands, given that the {\it U} filter
is highly affected by CN molecular bands.


\section{Multimodal MSs. The cases of NGC 2808 and NGC 6752}
\label{NGC2808}
  In few cases the MS morphology strongly supports the presence of
  stellar generations with different helium content. 
%
The most striking evidence comes from 
the color magnitude diagram (CMD) of NGC 2808 which exhibits two additional 
MSs that run blueward of the main MS ridge line  
 and have been associated to subsequent episodes of star formation (Piotto
et al.\ 2007).

Apart from the triple MS, NGC 2808 shows observational
evidence indicating the presence of multiple stellar populations also
in other region of the CMD.
Its horizontal branch (HB) is greatly extended bleward and is well populated 
on both sides of the instability strip. The
distribution of stars along the HB is multimodal (Sosin et al.\ 1997,
Bedin et al.\ 2000) with three significant gaps, one of these gaps
being at the color of the RR Lyrae instability strip. In fact, even 
though the HB is well populated both to the blue and to the red of the
instability strip, very few RR Lyrae 
stars have been identified in NGC 2808. The other two gaps are on the
blue extension of the HB and delimit three distinct segments. 
      
Even the RGB is not consistent with a single stellar
population. The CMDs shown by Yong et al.\ (2007) and Lee et
al.\ (2009) revealed a large color spread among RGB stars that cannot
be attributed to photometric errors only.
Furthermore, an analysis of medium-high-resolution spectra of 122 RGB stars
have revealed an extended Na-O anticorrelation in NGC 2808 (Carretta
et al.\ 2006) with the presence of three distinct groups:
O-normal (peak at
[\rm O/Fe] $+0.28$), O-poor (peak at [\rm O/Fe] $-0.21$)
and super-O-poor (peak at [\rm O/Fe] $-0.73$) stars. 

On the basis of their relative numbers, Piotto et al.\ (2007)
associated the three MSs with the three HB segments defined by Bedin
et al.\ (2000), and the three groups of stars with different O content found by
Carretta et al.\ (2006). 
Since cluster stars have around the same iron content (Carretta et
al.\ 2007, 2010), a  multimodal distribution of He abundances seems to
be the only way to take into account for both the complex HB
and the multiple MS (D'Antona et al.\ 2005) and is consistent with the
observed abundance pattern. In this case the
population associated to the red MS (rMS) have nearly primordial helium,
while stars of the middle (mMS) and blue MS (bMS) 
are formed from the ejecta produced by an earlier stellar generation
through the complete CNO and MgAl cycle and are He-enhanced (Y$\sim$0.33 and
Y$\sim$0.38, respectively).  
This scenario is nicely confirmed by the recent work of Bragaglia et
al.\ (2010) who measured  chemical abundances of
one star on the rMS and one ond the bMS and
 found that the latter shows an enhancement of N, Na, and Al and a depletion
of C and Mg as expected for material polluted by first-generation
massive stars.

\begin{figure}[ht!]
 \centering
 \includegraphics[width=0.85\textwidth]{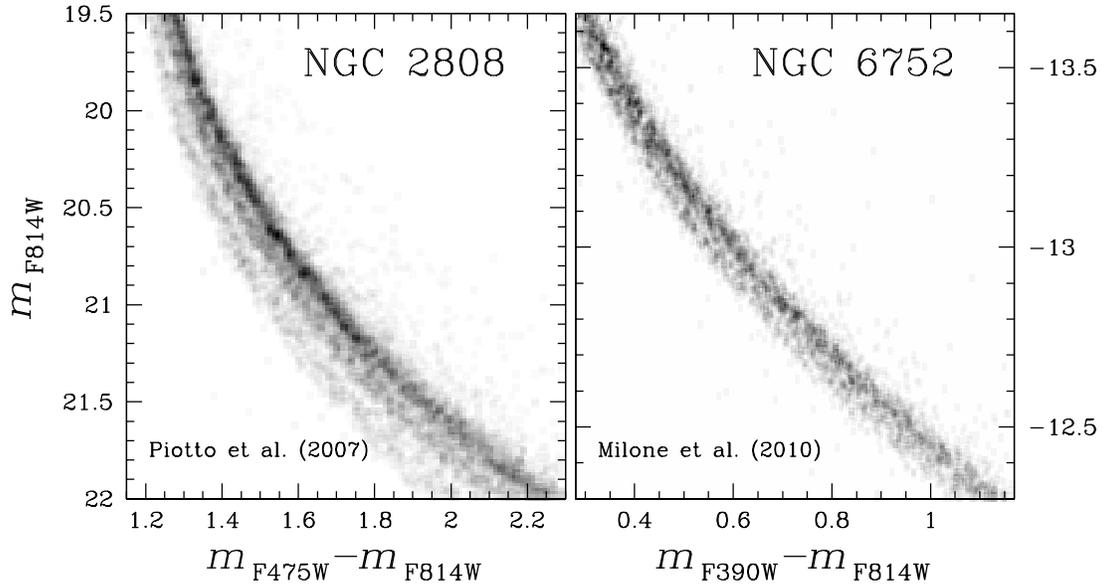}      
  \caption{Hess diagram of NGC 2808 and NGC 6752 zoomed around the MS region.} 
  \label{fig:ms}
\end{figure}

A split or a broad MS is not a peculiarity of NGC
2808 but is present also in other GCs like 47 
Tucanae and NGC 6752 (Anderson et al. \ 2009, Milone et
al. 2010).  In these cases the color spread
has been tentatively attributed to small variations in He
($\Delta$ Y$\sim$0.02-0.03). As an example in Fig.~\ref{fig:ms} we show the
Hess diagram for MS stars of NGC 2808 and NGC 6752 from Hubble space
telescope {\it HST} ACS/WFC and WFC3/UVIS cameras.

\section{SGB split clusters: NGC 1851 and NGC 6656}
\label{NGC1851-6656}  
\begin{figure}[ht!]
 \centering
 \includegraphics[width=\textwidth]{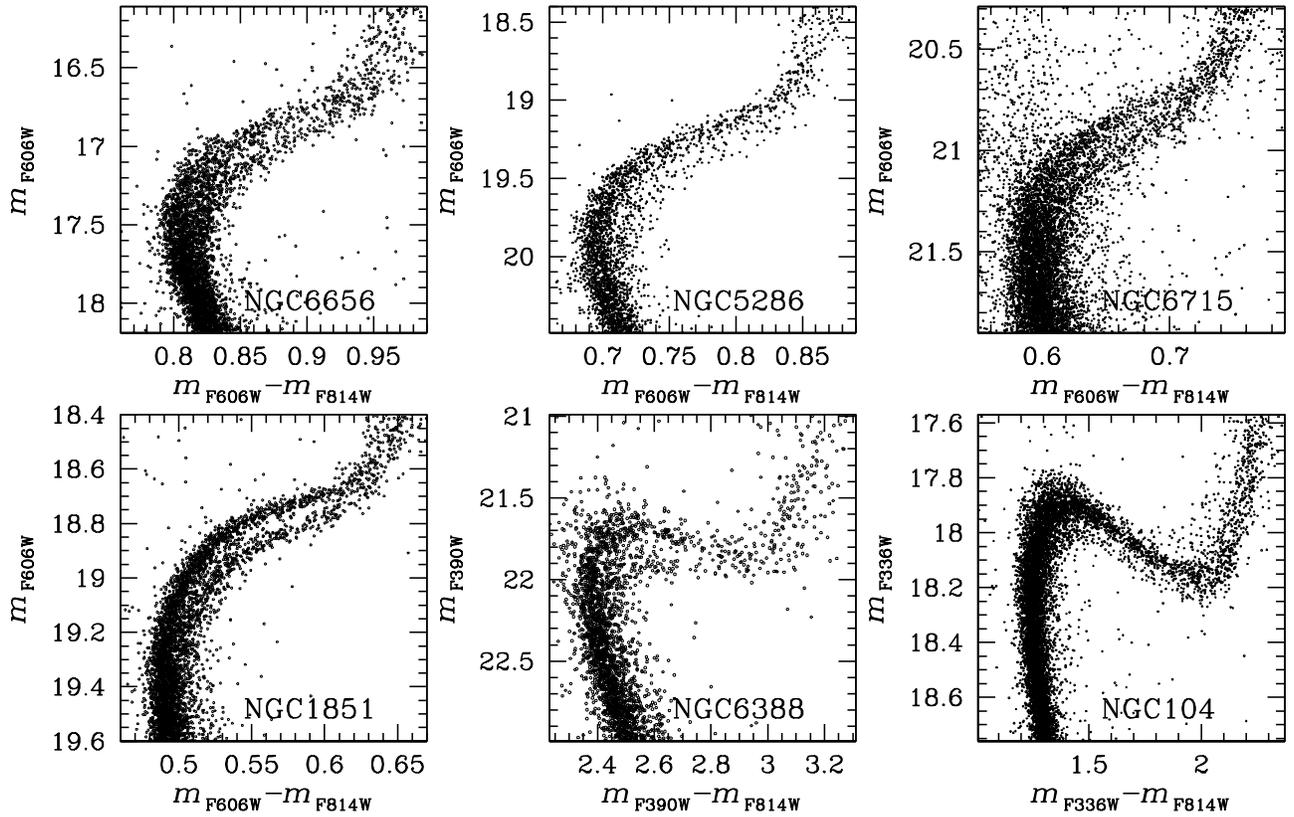}      
  \caption{Collection of CMDs from {\it HST} ACS/WFC and WFC3/UVIS data for
    six GCs with spread or split SGB. (Piotto et.\ 2010, Milone et
   al.\ 2008, 2010, Anderson et al.\ 2009). } 
  \label{fig:sgb}
\end{figure}

High-accuracy photometry obtained with the WFPC2, the WFC/ACS and the
WFC3/UVIS cameras on
board of the {\it HST} has revealed that in several clusters there is a
broad, split, or multimodal SGB as shown in Fig.~\ref{fig:sgb} for NGC
1851, NGC 6388, NGC 104, NGC   6656, NGC 5286, and NGC 6715  (Milone
et al.\ 2008, Anderson et al.\ 2009, Piotto et al.\ 2009, Moretti et al.\ 2008). 

NGC 1851 and NGC 6656 are the most studied clusters of this
 group and will be analyzed in more detail in the following.
We find that
the SGB of both NGC 1851 and NGC 6656 is clearly split into two branches
with the bright SGB 
component containing  about the 60\% of the total number of SGB
stars and the remaining 40\% of stars belonging to the faint SGB
(Milone et al.\ 2008, Piotto \ 2009, Marino et al.\ 2009). 

These results have brought new interest on these GCs and a lot of
effort have been made to understand their star formation history.
Theoretic studies demonstrate that the double SGB can been explained in
terms of two stellar groups,  
only slightly differing in age, with the younger one having an increased
C+N+O abundance (Cassisi et al.\ 2008, Ventura et al.\ 2009).
Indeed, significant star-to-star variations in the overall CNO
abundance have been detected  among NGC 6656 RGB stars by Marino et
al.\ (2010) and in two out four NGC 1851 giants by Yong et al. \ (2009).
 As an alternative possibility, we note that the double SGB is consistent with
two stellar populations  with constant CNO but differing in age by
$\sim$1 Gyr (Milone et al.\ 2008).  

\begin{figure}[ht!]
 \centering
 \includegraphics[width=0.75\textwidth]{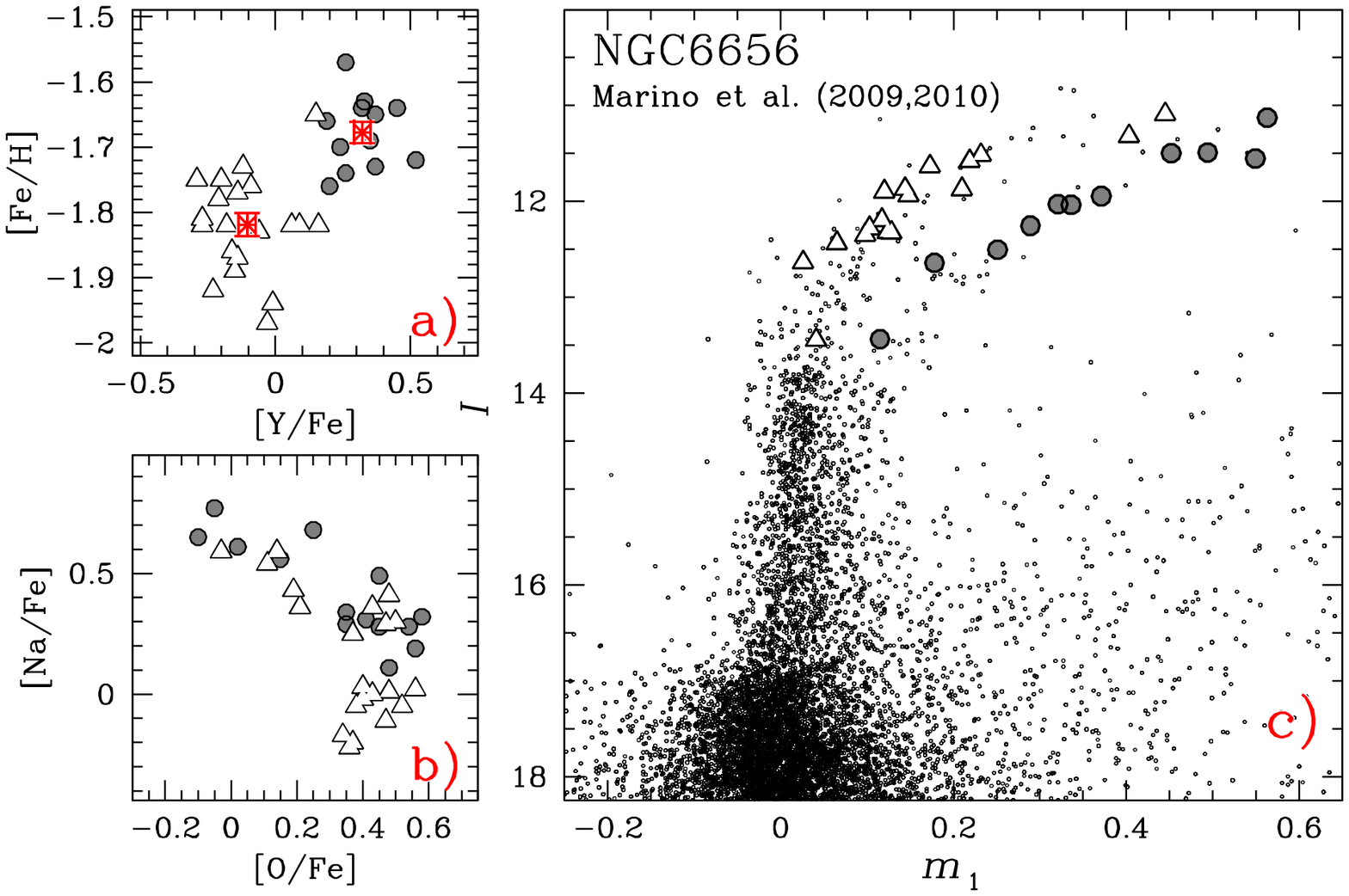}      
  \caption{{\it Panel a)}: [\rm Fe/H] as a function of [\rm Y/Fe] for
    RGB stars in NGC 6656. The two groups of $s$-rich and $s$-poor
    stars are plotted as gray circles and white triangles,
    respectively. The Na-O anticorrelation for 
    the same stars of NGC 6656 is shown in {\it Panel b)} while in {\it Panel
      c)} we marked them in the {\it I} versus  ${\it m}_{\rm
      1}$ diagram from Ritcher et al.\ (1999). } 
  \label{M22}
\end{figure}

A peculiar property of the cluster pair NGC 1851-NGC 6656 is the large scatter 
 in the abundance of those $n$-capture elements that are associated to
 $s$-process (Marino et al.\ 2009, 2010, Hesser et  al.\ 1982,
 Yong et al.\ 2008). 
In both clusters $n$-capture elements are clearly segregated around
two distinct values of barium and yttrium in sharp contrast with what
found in most GCs, where the 
abundance of these elements does not exhibit significant star-to-star
variations. 
As an example, Fig.~\ref{M22}a  shows the iron abundance as a
function of [\rm Y/Fe] for NGC 6656.  

These recent results nicely match with previous photometric analysis by
Ritcher et al.\ (1999) and Calamida et al.\ (2007) who found that NGC 6566
and NGC 1851 exhibit a bimodal distribution in the ${\it m}_{1}$ index
among RGB stars. A similar RGB bimodality has
been observed in the {\it hk} index by Lee et al.\ (2009) and in the
{\it U} vs. ({\it U$-$I}) and {\it U} vs. ({\it U$-$V}) CMD by Han et
al.\ (2009) and Momany et al.\ (2004). 
In these cases the RGB components are clearly associated to the two SGBs. 

In the light of these results we matched spectroscopic data of NGC
6656 from Marino et al.\ (2009, 2010) with the Stro\"mgren photometry by
Richter et al.\ (1999). Results are shown in
 Fig.~\ref{M22}c where we plotted the
$m_{1}$ versus {\it I} diagram for NGC 6656 corrected for differential
reddening and found that $s$-rich and $s$-poor stars define two
distinct RGBs. 
As the $m_{1}$ index is strongly dependent by the CN bands strength,
we expect this bimodality as due to the overabundance in both C and N
measured by Marino et al.\ (2010) in $s$-rich stars. 

A fundamental piece of the puzzle comes from the [\rm Fe/H] measurements.
While the presence of a possible iron dispersion among stars in NGC
1851 is still controversial (Yong et al.\ 2008, Carretta et al.\ 2010,
Villanova et al.\ 2010), when we compare the iron abundance for NGC
6656 with the $s$-elements 
abundance, we find a strong correlation, with $s$-rich stars having
a systematically higher [Fe/H] of $\delta {\rm [Fe/H]}=0.14\pm0.03$ (Marino et
al.\ 2009) as shown in Fig.~\ref{M22}a.
 This result demonstrates that, at odds with `normal'
monometallic GCs the different stellar populations of NGC 6656 have
significant differences in their iron content and that core-collapse supernovae 
played a prominent role in the star formation history of this cluster. 

An intriguing property of this pair of clusters 
is that both the $s$-rich and the $s$-poor 
group of stars have its own Na-O anticorrelation as shown in
Fig~\ref{M22}b for NGC 6656.   
For the latter a C-N anticorrelation has also been detected
in both the $s$-groups by Marino et al.\ (2010).
Variations in light elements have been considered as the
signature of multiple stellar populations, therefore this result may
indicate that
 NGC 1851 and NGC 6656 have experienced a very complex star formation
 history with the presence of two stellar groups with discrete
 $s$-elements abundance, each containing multiple generation of stars
 with different sodium and oxygen content.





\section{Conclusions}
For several decades GCs have been considered as the best approximation
of simple stellar populations consisting of coeval and chemically
homogeneous stars. This picture have been mainly challenged by two 
observational facts. 

Since the seventies we know that GCs exhibit a peculiar pattern in
their chemical abundances with large star-to-star variations in the
abundances of C, N, Na, O, Mg, and Al. These variations are primordial
since they are observed in stars at all the evolutionary phases and
are peculiar to GC stars. Field stars only changes in C and N
abundance expected from typical evolution of low-mass stars.  

In addition, since the early sixties we know that the HB of some GCs
are quite peculiar. The distribution of stars along the RGB can be
multimodal with the presence of one or more gaps and in same cases the
HB can be extended toward very high temperatures.
It is well known metallicity is the first parameter governing the
HB morphology there are some GCs with almost the same iron content but
different HB morphology demonstrating the metallicity alone is not
enough to reproduce the observational scenario.
This problem, known as the {\it second-parameter} problem, still lacks
of a comprehensive understanding.

 The recent discoveries of multiple stellar populations in GCs 
have shattered once and for all the long-held paradigm of GCs as simple
stellar populations and brought new interest on these stellar systems.

It is very tempting to relate the second parameter HB problem to the
complex abundance  pattern of GCs as well as to the multiple sequences
observed in the CMD of some clusters. As already mentioned the observed
variations of light elements indicate the presence of material
processed through hot H-burning processes and should be also
He-enriched. While small variations in helium content should have a
small impact on colors and magnitudes for MS stars a large impact is
expected on the colors of HB stars since He-rich stars should be also
less massive. 

In summary the discovery of multiple stellar populations started a
new era on globular  cluster research.
While the observational scenario is still puzzling and there is
 a rather incoherent picture of the multipopulation phenomenon,
for the first time we  
might have the key to solve a number of problems, like the abundance
anomalies and possibly the second parameter problem, as well as the
newly discovered multiple sequences in the CMD.

\begin{acknowledgements}
We gratefully acknowledge M. Hilker, R. Kraft,  C. Sneden and
G. Wallerstein for sending us their data. We wish to warmly tank Ivan
R.\ King and Jay Anderson, without whom most of the results presented in
this review would not have been possible. A special thanks to
A.\ Bragaglia, E.\ Carretta, S.\ Cassisi, M. Catelan,  F.\ D'Antona, M.\ Di
Criscienzo, R.\ Gratton, S. Lucatello, A.\ Moretti, A.\ Renzini, M. Salaris,
P.\ B. Stetson, P. Ventura, S.\ Villanova, M.\ Zoccali, A.\ Sarajedini and
the HST/ACS GCs Tresaury group for the many discussions on
multipopulations in Globular Clusters. 
\end{acknowledgements}

%
%
%
%
%

%
\end{document}